\documentclass[runningheads]{svmult}

\usepackage{makeidx}   
\usepackage{graphicx}  
\usepackage{subeqnar}  
\usepackage{multicol}  
\usepackage{physprbb}  
\makeindex             

\newcommand{\ie}{{i.e.,}\ }
\newcommand{\etal}{{\it et~al.\ }}

\newcommand{\lesssimrc}{\ {\hskip-4pt\raise-.5ex\hbox{$\buildrel<\over\sim$}}\ }
\newcommand{\gtrsimrc}{\ {\hskip-4pt\raise-.5ex\hbox{$\buildrel>\over\sim$}}\ }

\begin{document}
\title*{The Planetary Nebula Luminosity Function}
\toctitle{The Planetary Nebula Luminosity Function}
\titlerunning{The Planetary Nebula Luminosity Function}
%
\author{Robin Ciardullo}
\authorrunning{Robin Ciardullo}
%
%
\institute{The Pennsylvania State University,
           Department of Astronomy and Astrophysics
           525 Davey Lab
           University Park, PA 16803, USA}

\maketitle              

\begin{abstract}
The [O~III] $\lambda 5007$ planetary nebula luminosity function (PNLF) occupies
an important place on the extragalactic distance ladder: it is the only
standard candle that can be applied to all the large galaxies of the Local
Supercluster.  We review the method's precision, and use it to show that 
the distance scale defined by Cepheids and the Surface Brightness Fluctuation
method is likely too large by $\sim 7\%$.   We also
discuss some of the physics underlying the phenomenon, and present clues
as why the technique is so resilient.

\end{abstract}

\section{Introduction}
Planetary nebulae (PNe) are unique probes of galactic chemical and dynamical
evolution.  But before they can be used, they must be found, and the
best way of doing this involves searching for PN candidates via their bright 
[O~III] $\lambda 5007$ emission.  As a result,
the first piece of information obtained about a set of extragalactic
PNe is their [O~III] $\lambda 5007$ luminosity function.  Since this 
luminosity function contains information about both distance and parent 
population, one should make every effort to understand its properties and 
features.

\section{The PN Luminosity Function as a Distance Indicator}

\begin{figure}[t]
\begin{center}
\includegraphics[width=1.0\textwidth]{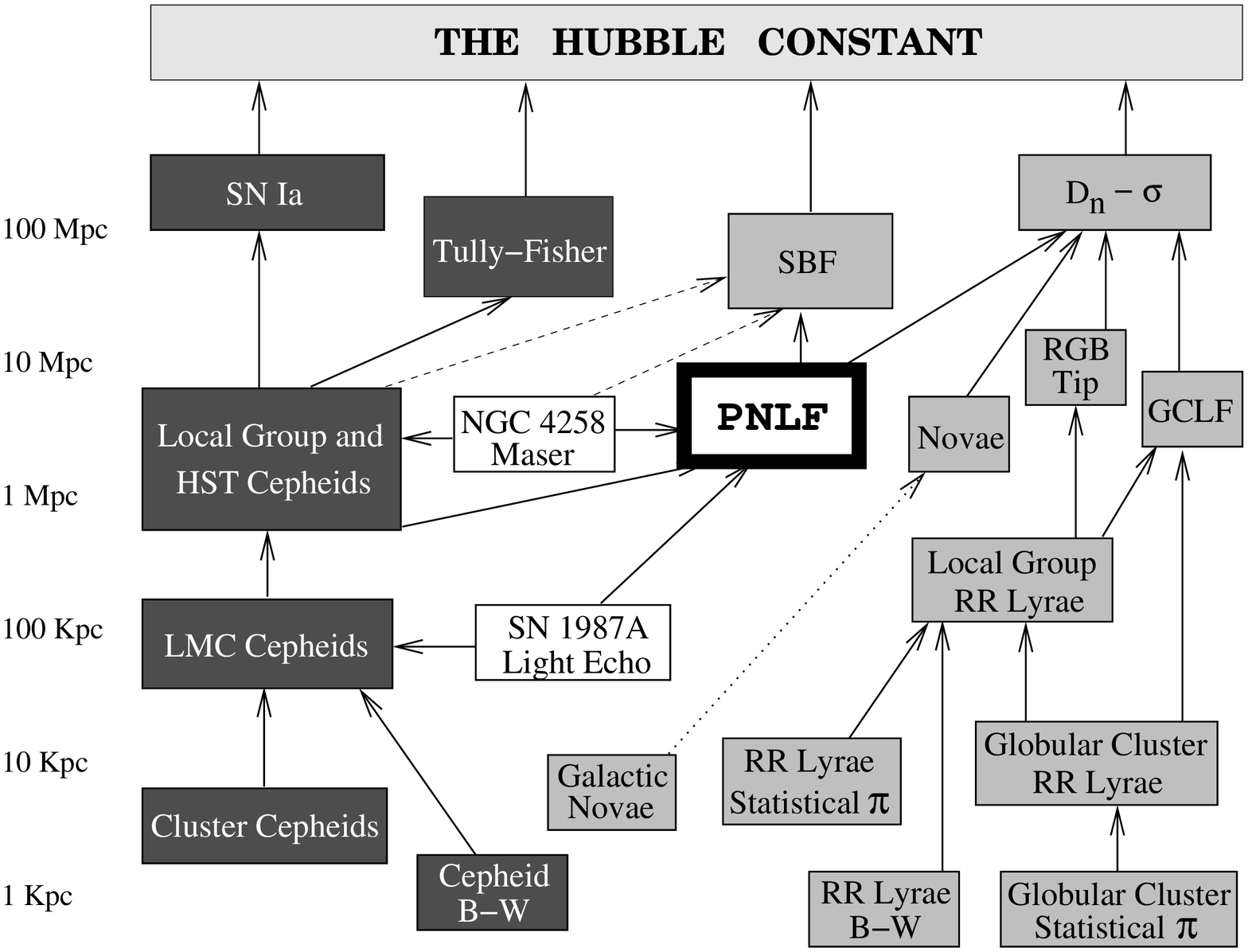}
\end{center}
\caption[]{The extragalactic distance ladder.  The dark boxes show 
techniques useful in star-forming galaxies, the lightly-filled boxes give 
methods that work in Pop~II systems, and the open boxes represent 
geometric distance determinations.  Uncertain calibrations are noted as
dashed lines.  The PNLF is the only method that is equally effective in all
the populations of the Local Supercluster.}
\label{eps1}
\end{figure}

Although the idea of using PNe as extragalactic distance indicators was
first suggested the early 1960's \cite{hw63,hodge}, it was not until the late 
1970's that a PN-based distance estimate was made \cite{fj78}, and the first
study of the [O~III] $\lambda 5007$ planetary nebula luminosity function (PNLF) 
was not performed until 1989 \cite{p1,p2}.  In fact, it is an irony of the 
subject that PN distance measurements inside the Local Group \cite{jl81} 
were made only after the technique had been applied to more distant systems
\cite{fj78,lg83}, and the first application of the PNLF inside the Milky Way
\cite{pottasch} occurred after the method had been used to estimate the 
Hubble Constant \cite{p5}.  This odd chronology, of course, was due to the fact
that individual PNe are definitely not standard candles 
\cite{berman,minkowski,shklovsky}, and that distance estimates to Galactic
PNe are extremely poor.  Nevertheless, today the [O~III]
$\lambda 5007$ PNLF is one of the most important standard candles in 
extragalactic astronomy, and the only method that can be applied to 
all the large galaxies of the Local Supercluster, regardless of environment or 
Hubble type (see Fig.~\ref{eps1}).

The use of the PN luminosity function is extremely simple.  One surveys a
distant galaxy or cluster and identifies point sources that are present
in [O~III] $\lambda 5007$ but completely invisible in the continuum.
These are the planetary nebula candidates.  Unfortunately, mixed in with the
PNe are three types of contaminants:

{\it H~II Regions:}  Star formation is not an issue in most elliptical
and lenticular galaxies, but in late-type systems, the H~II regions around
O and B stars will far outnumber the PNe.  Fortunately, most H~II regions are
resolvable under good seeing ($\lesssimrc 1''$) conditions, and their 
exciting stars may be visible on deep continuum images.  This contrasts with 
PNe which (in galaxies more distant than the Magellanic Clouds) are always 
stellar and have invisible central stars.  Moreover, observations in M31's 
bulge, M33's disk, and the LMC have shown that all planetaries in the top 
$\sim 1$~mag of the PNLF have [O~III] $\lambda 5007$ to H$\alpha$+[N~II] line 
ratios greater than $\sim 1.6$ \cite{p12}.  Most H~II regions have H$\alpha$ 
brighter than [O~III] \cite{shaver}.    

{\it Supernova Remnants:}  Unresolved high-excitation supernova remnants
can masquerade as planetary nebulae, especially in galaxies that have a cold, 
high-density interstellar medium.  Because compact supernova remnants are rare,
their effect on the PNLF is minimal.  Nevertheless, one needs to be aware
of this source of contamination, especially at the bright limit of the 
luminosity function.  Any single object that appears overluminous in 
[O~III] $\lambda 5007$ could be the result of a supernova explosion.

{\it Ly$\alpha$ Galaxies:}  At $z = 3.12$, Ly$\alpha$ is redshifted to
5007~\AA, and at fluxes below $\sim 10^{-16}$~ergs~cm$^{-2}$~s$^{-1}$,
unresolved and marginally resolved high-redshift galaxies with extremely 
strong Ly$\alpha$ emission (equivalent widths $\gtrsimrc 300$~\AA\ in the 
observers frame) can mimic planetary nebulae \cite{kudritzki,freeman}.  Since 
the surface density of these starbursting objects is relatively low, 
$\sim 1$~arcmin$^{-2}$~per unit redshift interval brighter than $5 \times 
10^{-17}$~ergs~cm$^{-2}$~s$^{-1}$ \cite{blank}, PN surveys in galaxies are 
not strongly affected by this contaminant.  However, in the intracluster 
environment of systems such as Virgo and Fornax, Ly$\alpha$ galaxies are the 
predominant source of error, and they limit the effectiveness of any luminosity 
function analysis.  To identify these interlopers, one must either obtain 
deep broadband images (to detect the underlying continuum of the galaxies)
or perform spectroscopy (to test for the existence of [O~III] $\lambda 4959$ 
or resolve the $\sim 400$~km~s$^{-1}$ width of the Ly$\alpha$ line).

Once the PNe are found, all one needs to do to determine a distance is to
measure their monochromatic [O~III] $\lambda 5007$ fluxes, define a 
statistical sample of objects, and fit the observed luminosity function to 
some standard law.  For simplicity, Ciardullo \etal \cite{p2} have described 
the PNLF via a truncated exponential
\begin{equation}
N(M) \propto e^{0.307 M} \{ 1 - e^{3 (M^* - M)} \}
\end{equation}
where
\begin{equation}
m_{5007} = -2.5 \log F_{5007} - 13.74
\end{equation}
though other forms of the relation are possible \cite{men93}.  In the above
equation, the key parameter is $M^*$, the absolute magnitude of the brightest
planetary. Despite some efforts at Galactic calibrations \cite{pottasch,men93},
the PNLF remains a secondary standard candle.  The original value for the zero 
point, $M^* = -4.48$, was based on an M31 infrared Cepheid distance of 
710~kpc \cite{welch} and a foreground extinction of $E(B-V)=0.11$ 
\cite{mcclure}.  Since then, M31's distance has increased \cite{keyfinal}, 
its reddening has decreased \cite{schlegel}, and, most importantly, the 
Cepheid distances to 12 additional galaxies have been included in the 
calibration \cite{p12}.  Somewhat fortuitously, the current value of $M^*$
is only 0.01~mag fainter than the original value, $M^* = -4.47 \pm 0.05$.

Note that all PNLF distances require some estimate of the foreground
interstellar extinction.  There are two sources to consider.  The first,
extinction in the Milky Way, is readily available from reddening maps
derived from H~I measurements and galaxy counts \cite{bh84} and/or from the
DIRBE and IRAS satellite experiments \cite{schlegel}.  The second, internal
extinction in the host galaxy, is only a problem in late-type spiral and
irregular galaxies.  Unfortunately, it is difficult to quantify.
In the Galaxy, the scale height of PNe is significantly larger
than that of the dust \cite{mb81}.  If the same is true in other galaxies,
then we would expect the bright end of the PNLF to always be dominated 
by objects foreground to the dust layer.  This conclusion seems to be supported
by observational data \cite{p11,p12} and numerical models \cite{p11}, both
of which suggest that the internal extinction which affects a galaxy's PN 
population is $\lesssimrc 0.05$~mag.  We will revisit this issue in
Section~3.

\section{Tests of the Technique}
In the past decade, the PNLF has been subjected to a number of rigorous
tests.  These tests fall into four categories.

\subsection{Internal Tests Within Galaxies}  Five galaxies have large enough 
PN samples to test for radial gradients in the location of the PNLF cutoff.
Two are large-bulge Sb spirals (M31 \cite{hui94} and M81 \cite{mag81}), one
is a pure-disk Sc spiral with a strong metallicity gradient (M33 \cite{m33}), 
one is large elliptical galaxy (NGC~4494 \cite{p10}), and one is a blue, 
interacting peculiar elliptical (NGC~5128 \cite{hui93}).  No significant 
change in the PNLF has been seen in {\it any\/} of these systems.  Given the 
diversity of the stellar populations sampled in these systems, this result, in 
itself, is impressive proof of the robustness of the method.

\subsection{Internal Tests Within Galaxy Groups}  To date, six galaxy clusters
have multiple PNLF measurements:  the M81 Group (M81 and NGC~2403 
\cite{p3,p11}), the NGC~1023 Group (NGC~891 and 1023 \cite{p7}), the NGC~5128 
Group (NGC~5102, 5128, and 5253 \cite{mcmillan,hui93,n5253}), the Fornax 
Cluster (NGC~1316, 1380, 1399, and 1404 \cite{p9}), the Leo~I Group (NGC~3351, 
3368, 3377, 3379, and 3384 \cite{p12,p11,p4}), and the Virgo Cluster 
(NGC~4374, 4382, 4406, 4472, 4486, and 4649 \cite{p5}).  In each system, the 
observed galaxies have a range of color, absolute magnitude, and Hubble type.  
Despite these differences, no discrepant distances have been found, as the 
galaxies of each group always fall within $\sim 1$~Mpc of each other.  Indeed, 
the PNLF measurements in Virgo easily resolve the infalling M84/M86 Group, 
which is background to the main body of the cluster \cite{bohringer}.

\subsection{External Tests with Different Methods}
Because planetary nebulae are found in all types of galaxies, it is possible
to compare the results of the PNLF method directly with distances derived from 
all of the other techniques at the top of the distance ladder.  The most 
instructive of these comparisons involve Cepheid variables and the Surface 
Brightness Fluctuation (SBF) Method.

\begin{figure}[t]
\begin{center}
\includegraphics[width=1.0\textwidth]{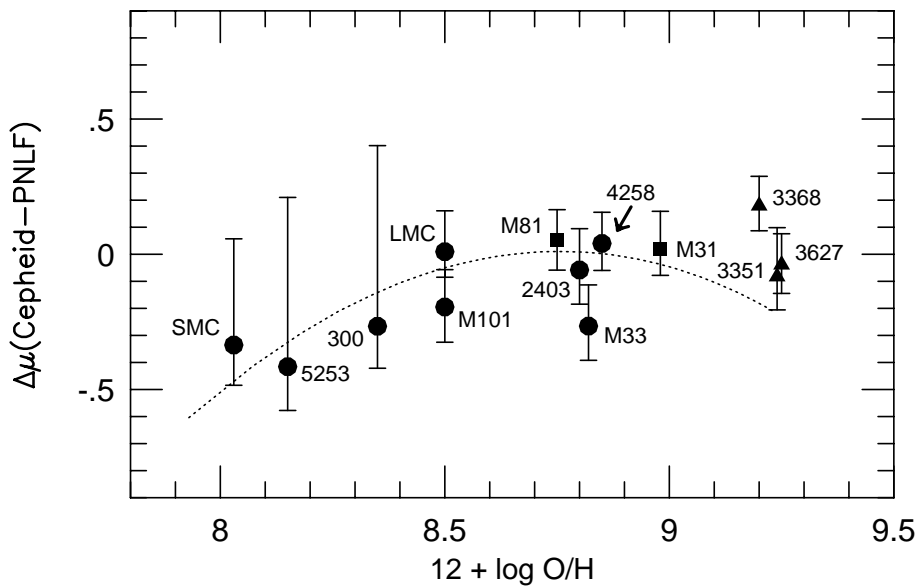}
\end{center}
\caption[]{A comparison of the PNLF and Cepheid distance moduli as function
of galactic oxygen abundance, as estimated from the systems' H~II regions 
\cite{fdatabase}.  No metallicity correction has been applied to either
distance indicator.  Circles represent PNLF measurements in galactic inner 
disks, squares in galactic bulges, and triangles in outer disks/halos.  
The error bars display the formal uncertainties of the methods added in
quadrature; small galaxies with few PNe have generally larger errors.
The curve shows the expected reaction of the PNLF to metallicity \cite{djv}.  
Note that metal-rich galaxies should not follow this relation, since these 
objects always contain enough lower metallicity stars to populate the PNLF's 
bright-end cutoff.  The scatter between the measurements is consistent with the
internal errors of the methods.}
\label{eps2}
\end{figure}

Figure~\ref{eps2} compares the PNLF distances of 13 galaxies (derived
using the foreground extinction estimates from DIRBE/IRAS \cite{schlegel})
with the final Cepheid distances produced by the {\sl HST Key Project\/}
\cite{keyfinal}.  Neither dataset has been corrected for the
effects of metallicity.   Since the absolute magnitude of the PNLF cutoff,
$M^*$, is calibrated via these Cepheid distances, the weighted mean of the 
distribution must, by definition, be zero.  However, the scatter about this 
mean, and the systematic trends in the data, are valid indicators of the 
reliability of the technique.

As Fig.~\ref{eps2} illustrates, the scatter between the Cepheid 
and PNLF distances is impressively small.  Except for the most 
metal-poor systems, the residuals are perfectly consistent with the internal 
uncertainties of the methods.  Moreover, the systematic shift seen at 
low-metallicity is exactly that predicted by models of post-AGB evolution
\cite{djv}.  There is little room for any additional metallicity term, 
either in the Cepheid or PNLF relations.

\begin{figure}[t]
\begin{center}
\includegraphics[width=1.0\textwidth]{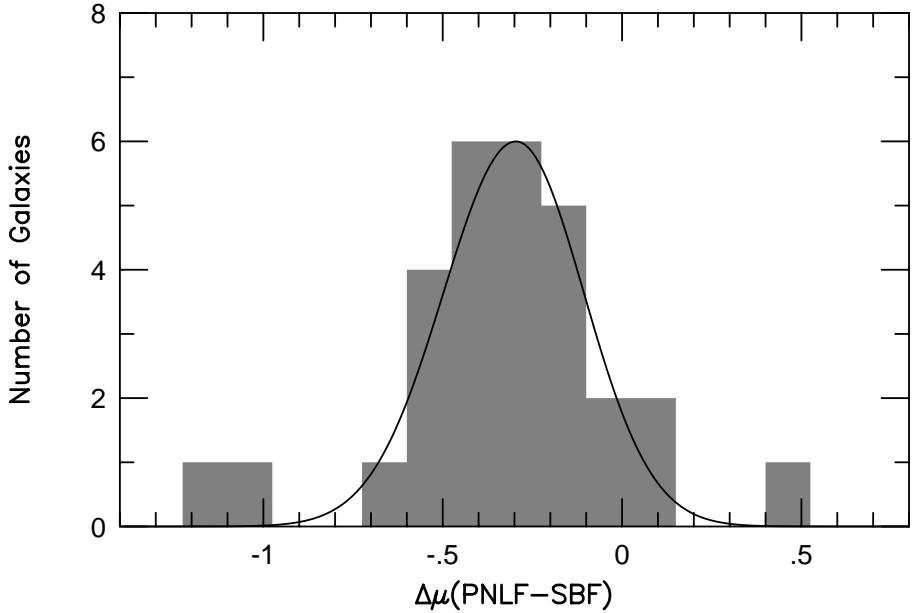}
\end{center}
\caption[]{A histogram of the difference between the PNLF and SBF
distance moduli for 29 galaxies measured by both methods.  The two worst
outliers are the edge-on galaxies NGC~4565 ($\Delta\mu = -0.80$) and
NGC~891 ($\Delta\mu = +0.71$).  The curve represents the expected dispersion 
of the data.  The figure demonstrates that the absolute scales of the two 
techniques are discrepant by $\sim 0.3$~mag, but the internal and external 
errors of the methods agree.}
\label{eps3}
\end{figure}

Figure~\ref{eps3} compares the PNLF distances of 29 galaxies with distances 
derived from SBF measurements.  As is illustrated, the relative distances of 
the two methods are in excellent agreement:  the scatter in the data is exactly
that predicted from the internal errors of the measurements.  However, the
absolute distance scale derived from the SBF data is $\sim 0.3$~mag larger
than that inferred from the PNLF{}.  Since both techniques are calibrated 
via Cepheid measurements, and both have formal zero-point uncertainties of 
$\sim 0.05$~mag, this type of offset should not exist.

The most likely cause of the discrepancy is a small amount of internal 
extinction in the Cepheid calibration galaxies.  For most methods (including
the PNLF), an underestimate of extinction produces an underestimate of 
luminosity.  As a result, undetected internal extinction within a Cepheid 
calibration galaxy propagates into an underestimate of the standard candle's 
brightness, and an extragalactic distance scale that is too small.  For the
SBF technique, however, the opposite is true.  The SBF standard candle,
$\bar M_I$, is extremely sensitive to color, \ie
\begin{equation}
\bar M_I = -1.74 + 4.5 [ (V-I)_0 - 1.15]
\end{equation}
\cite{tonry}.  If internal extinction exists in an SBF calibrator, the galaxy's
underlying population will appear too red, and the inferred value of 
$\bar M_I$ will be too bright.  The result will be a set of distances that
are systematically too large.

If internal extinction really is responsible for the offset displayed in
Fig.~\ref{eps3}, then the zero points of both systems must be adjusted.  These 
corrections propagate all the way up the distance ladder.  For example, 
according to the {\sl HST Key Project,} the SBF-based Hubble Constant is $69 
\pm 4$ (random) $\pm 6$ (systematic) km~s$^{-1}$~Mpc$^{-1}$ \cite{keyfinal}. 
However, if we assume that the calibration galaxies are internally reddened by 
$E(B-V) \sim 0.04$, then the zero point of the SBF system fades by 0.17~mag, 
and the SBF Hubble Constant increases to 75~km~s$^{-1}$~Mpc$^{-1}$.  This one 
correction is as large as the technique's entire systematic error budget.  
Such an error could not have been found without the cross-check provided by 
PNLF measurements.

\subsection{External Comparisons with Geometric Distances}Two galaxies have
distance measurements that do not depend on the distance ladder.
The first is NGC~4258, which has a resolved disk of cold gas orbiting 
its central black hole.  The proper motions and radial accelerations of water 
masers associated with this gas yield an unambiguous geometric distance of 
$7.2 \pm 0.3$~Mpc \cite{herrnstein}.  The second benchmark comes from the 
light echo of SN~1987A in the Large Magellanic Cloud.  Although the geometry 
of the light echo is still somewhat controversial, the most detailed and 
complete analysis of the object to date gives a distance of $D < 47.2 \pm 
0.1$~kpc \cite{gould}.  In Table~1 we compare these values with the distances 
determined from the PNLF \cite{p12} and from Cepheids 
\cite{keyfinal}.

\begin{table}
\caption{Benchmark Galaxy Distances}
\begin{center}
\setlength\tabcolsep{10pt}
\begin{tabular}{lccc}
\hline\noalign{\smallskip}
Method                &LMC    &NGC~4258 &$\Delta\mu$ (mag) \\
\hline
\noalign{\smallskip}
Geometry  &$< 18.37 \pm 0.04$ \phantom{$<$}
                                    &$29.29 \pm 0.09$  &$10.92 \pm 0.10$ \\
Cepheids  &18.50             &$29.44 \pm 0.07$  &$10.94 \pm 0.07$ \\
PNLF      &$18.47 \pm 0.11$  &$29.43 \pm 0.09$  &$10.96 \pm 0.14$ \\ 
\noalign{\smallskip}
\hline
\noalign{\smallskip}
\end{tabular}
\end{center}
\label{Tab1}
\end{table}

As the table demonstrates, the Cepheid and PNLF methods both overestimate
the distance to NGC~4258 by $\sim 0.14$~mag, \ie by $\sim 1.3 \, \sigma$ and 
$1.1 \, \sigma$, respectively.  In the absence of some systematic error 
affecting both methods, the probability of this happening is 
$\lesssimrc 5\%$.  On the other hand, there is no disagreement concerning
NGC~4258's distance {\it relative to that of the LMC:}  the Cepheids,
PNLF, and geometric techniques all agree to within $\pm 2\%$!  Such a
small error is probably fortuitous, but it does suggest the presence of
a systematic error that affects the entire extragalactic distance ladder.

In fact, the {\sl HST Key Project\/} distances are all based on an LMC
distance modulus of $(m-M)_0 = 18.50$ \cite{keyfinal}, and, via the data
of Fig.~\ref{eps2}, the PNLF distance scale is tied to that of the Cepheids. 
If the zero point of the Cepheid scale were shifted to $(m-M)_0 = 18.37$, then 
all the measurements would be in agreement.  This consistency supports a 
shorter distance to the LMC, and argues for a 7\% increase in the {\sl HST Key
Project\/} Hubble Constant to 77~km~s$^{-1}$~Mpc$^{-1}$.

\section{The Physics of the Luminosity Function}
The fact that the PNLF method is insensitive to metallicity is not a surprise.
Since oxygen is a primary nebular coolant, any decrease in its abundance
raises the plasma's electron temperature and increases the rate of collisional 
excitations per ion.  This mitigates the effect of having fewer emitting
ions in the nebula, so that a decrease in oxygen abundance only lessens the
emergent [O~III] $\lambda 5007$ line flux by (roughly) the square root of the 
abundance difference \cite{p1}.  Meanwhile, the PN's core reacts to metallicity
in the opposite manner.  If the metallicity of the progenitor star is
decreased, then the PN's central star will be slightly more massive and its
emergent UV flux will be slightly greater \cite{lattanzio,brocato}.  This 
additional energy almost exactly compensates for the decreased emissivity of 
the nebula.  As a result, the total [O~III] $\lambda 5007$ flux that is 
generated by a planetary is virtually independent of metallicity.  This 
fortuitous cancellation has been confirmed by more detailed models of PN 
evolution: only in the most metal-poor systems does the brightness of the PNLF 
cutoff fade by more than a few percent \cite{djv}.  

\begin{figure}
\begin{center}
\includegraphics[width=0.96\textwidth]{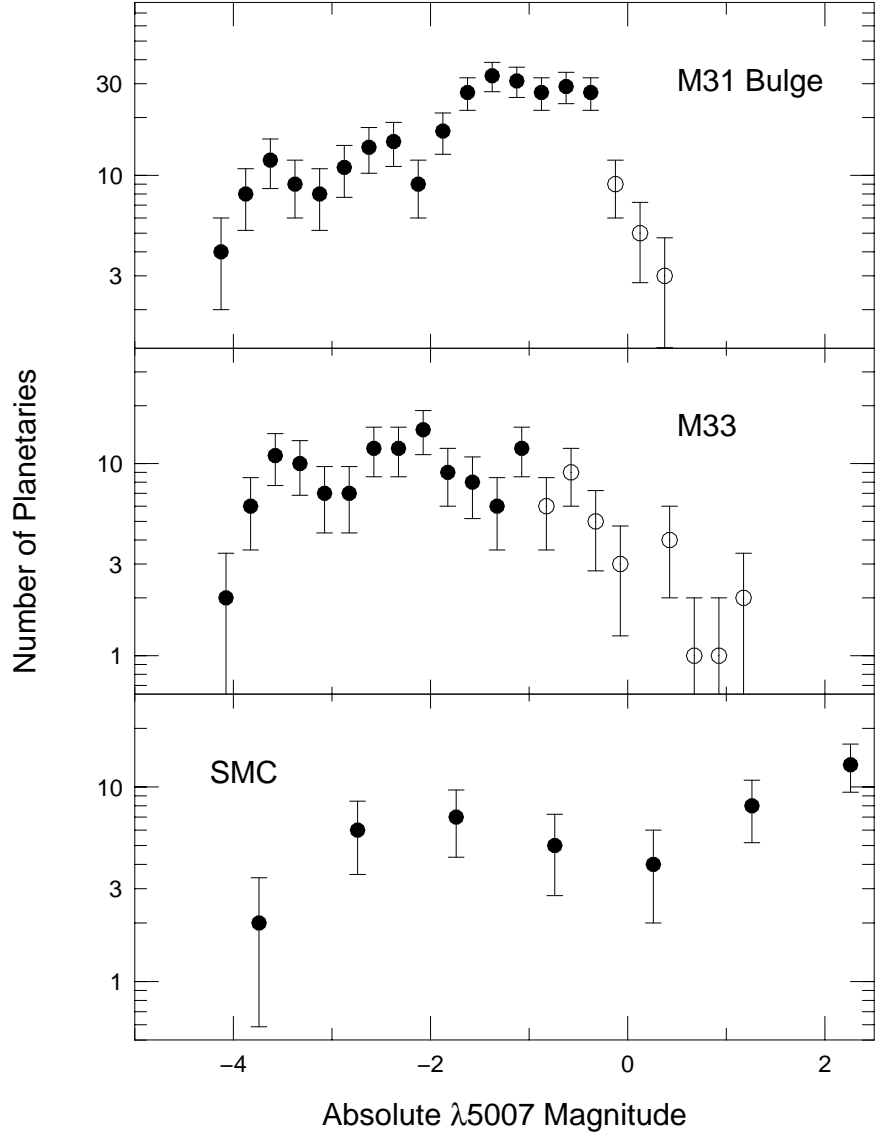}
\end{center}
\caption[]{The [O~III] $\lambda 5007$ planetary nebula luminosity functions of
of three Local Group stellar populations.  Old populations, such as that
found in the bulge of M31, have an exponentially increasing PNLF{}.  In
contrast, the PNLFs of star-forming systems are non-monotonic, with a
``dip'' that begins $\sim 2$~mag below $M^*$.  The location of this dip is 
consistent with models in which the luminosity evolution of planetary nebulae 
is governed by the rapid evolution of high-mass central stars.
} 
\label{eps4}
\end{figure}

The insensitivity of the PNLF cutoff to population age is more difficult to 
understand.  A PN's [O~III] $\lambda 5007$ flux is directly proportional to 
the luminosity of its central star, and this luminosity, in turn, is extremely 
sensitive to central star mass \cite{vw94,blocker}.  Since central star mass
is directly proportional to progenitor mass via the initial mass-final mass
relation \cite{weidemann}, one would think that the PNLF cutoff would fade
dramatically with time \cite{marigo}.  Yet observations demonstrate that this 
fading does not occur.

Although we do not as yet have a complete theory for the age-invariance of
the PNLF, a few clues are beginning to emerge.  The first concerns the faint
end of the luminosity function.  Although the location of the PNLF cutoff does
not change with stellar population, the same is not true for the function's
overall shape.  As Fig.~\ref{eps4} demonstrates, old populations, such as 
M31's bulge, have a PNLF that increases exponentially following $N(m) \propto 
e^{0.307 m}$.  This type of behavior is expected if the PNe of these systems 
have low-mass cores, so that the timescale for central star evolution is much 
longer than that for nebular expansion \cite{hw63,m33}.  In contrast, 
star-forming galaxies have a ``dip'' in the PNLF at magnitudes between 
$\sim 2$ and $\sim 4$~mag down from $M^*$.  This deficit is a natural 
characteristic of models in which the evolution of [O~III] $\lambda 5007$ is 
governed by the rapid evolution of a high-mass core.  In such systems, the 
non-monotonic PNLF is simply a reflection of the bimodal luminosity function 
expected from post-AGB stars \cite{vw94}.  

A second clue to the PNLF comes from the fortuitous correlation between 
the maximum UV emission achieved by a PN central star and the mass of its
envelope.   High luminosity PN cores have large circumstellar envelopes,
and, consequently, large amounts of circumstellar dust.   The extinction caused
by this dust can act to regulate the amount of [O~III] $\lambda 5007$ flux
generated by a planetary, so that the luminosity of high core-mass objects 
never exceeds $M^*$.  Support for this scenario comes from the fact that
nine Magellanic Cloud planetaries have intrinsic [O~III] $\lambda 5007$
magnitudes brighter than $M^*$, but {\it all\/} are self-extincted below 
the PNLF cutoff \cite{md1,md2,p6}.  In addition, there is observational 
evidence for a correlation between the amount of PN circumstellar 
extinction and central star mass for young, [O~III]-bright objects \cite{cj99}.

The final constraint on the physics of the PNLF comes from the number of 
PN produced by different stellar populations.  According to the theory of
stellar energy generation \cite{renzini},
all populations older than $\sim 1$~Gyr create $\sim 1$~PN per year per
$5 \times 10^{12} L_{\odot}$ (bolometric).  If this is true, then measurements
of the luminosity-specific number of PNe present in a galaxy can be immediately
translated into constraints on the lifetime of an [O~III]-bright nebula and
on the fraction of stars which go through the bright planetary nebula phase.

\begin{figure}[t]
\begin{center}
\includegraphics[width=0.96\textwidth]{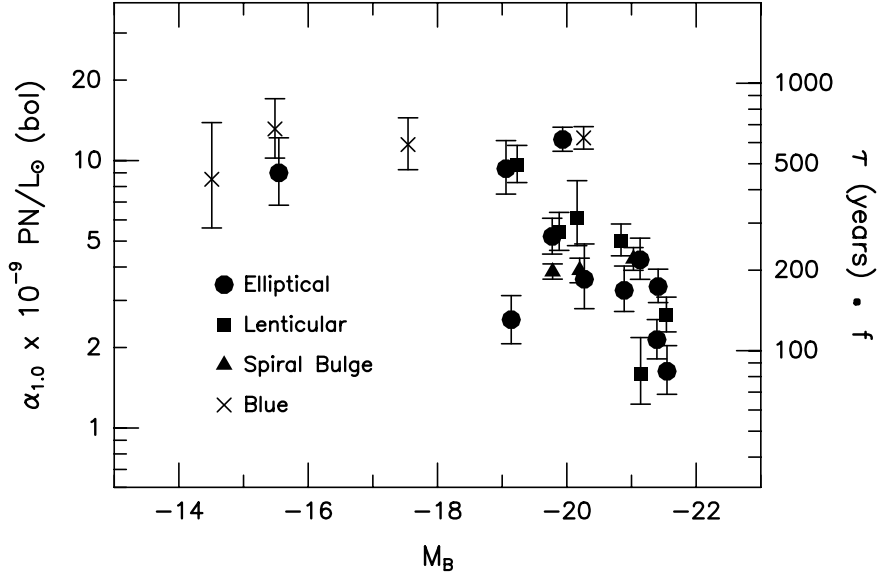}
\end{center}
\caption[]{The bolometric luminosity-specific number of PNe in the top
1~mag of the [O~III] $\lambda 5007$ PNLF plotted as a function of galactic
absolute $B$ magnitude.  The right side of the $y$-axis translates this
number into a constraint on the PN lifetime using a luminosity-specific 
stellar evolutionary flux of $2 \times 10^{-11}$~stars~yr${-1}~L_\odot$.  
Either the [O~III]-bright PNe of giant ellipticals live very short lives, or 
a substantial fraction of the stars in these galaxies do not contribute to the 
bright-end of the PNLF.}
\label{eps5}
\end{figure}

This constraint is given in Fig.~\ref{eps5}.  For small and medium-sized
galaxies, PNe typically spend $\sim 500$~yr within $\sim 1$~mag of $M^*$.
However, the apparent PN timescale decreases dramatically for objects in giant 
elliptical and lenticular galaxies.  Either the [O~III]-bright planetaries in 
these systems have much shorter lives than their small-galaxy counterparts, 
or a substantial fraction of the stars of giant ellipticals do not evolve into
these types of objects.

\section{The Future}
Of course, attempting to derive the evolution of ensembles of planetary nebulae
from [O~III] data alone is like trying to infer the evolution of star clusters
from photometry performed through only one filter.  In order to gain a fuller
understanding of the global properties of planetary nebulae, one must study
the luminosity function of PNe in a multi-dimensional emission-line space.
The data for such a study are just now beginning to be available.  Narrow-band 
photometric surveys are yielding information on the absolute line strengths of
PNe in [O~III] $\lambda 5007$ and H$\alpha$+[N~II] for a variety of stellar 
populations.  Follow-up radial velocity surveys are then producing data on
H$\beta$ relative to [O~III] $\lambda 5007$ and, in some cases, [N~II] and 
[S~II] relative to H$\alpha$.  By comparing these measurements to models, 
we should be able to gain a much more complete understanding of 
planetary nebula phenomenon.

\null\par
This work was supported in part by NSF grant AST 00-71238.

\null\par

%

\end{document}